\documentclass[a4paper,aps,onecolumn,nofootinbib]{revtex4}
\RequirePackage[colorlinks,hyperindex]{hyperref}
\RequirePackage[english]{babel}
\RequirePackage[latin1]{inputenc}
\RequirePackage[T1]{fontenc}
\RequirePackage{mathrsfs}
\RequirePackage{amsmath}
\RequirePackage{amssymb}
\RequirePackage{amsbsy}
\RequirePackage{color}
\RequirePackage{bm}

\hypersetup{colorlinks=true,breaklinks=true,urlcolor=blue,linkcolor=red}
\begin{document}
\title{\bf{Geometry of spinors: doubly-chiral plane-wave expansion}}
\author{Luca Fabbri$^{\nabla}$\!$^{\hbar}$\footnote{luca.fabbri@unige.it}}
\affiliation{$^{\nabla}$DIME, Universit\`{a} di Genova, Via all'Opera Pia 15, 16145 Genova, ITALY\\
$^{\hbar}$INFN, Sezione di Genova, Via Dodecaneso 33, 16146 Genova, ITALY}
\date{\today}
\begin{abstract}
We employ the polar re-formulation of spinor fields to see in a new light their classification into regular and singular spinors, these last also called flag-dipoles, further splitting into the sub-classes of dipoles and flagpoles: in particular, we will study the conditions under which flagpoles may be solutions of the Dirac field equations. We argue for an enlargement of the plane-wave expansion.
\end{abstract}
\maketitle
\section{Introduction}
The Lounesto classification of spinor fields \cite{L, Cavalcanti:2014wia, HoffdaSilva:2017waf, daSilva:2012wp, Ablamowicz:2014rpa}, built on the use of bi-linear spinorial quantities, entails an algebraic classifications of spinors according to whether some of these bi-linears are equal to zero or not. In $(1\!+\!3)$-dimensional space-times, and for $4\!\times\!4$ Clifford matrices, the aforementioned bi-linears are given by a scalar and a pseudo-scalar, a vector and an axial-vector, and a rank-$2$ antisymmetric tensor, respectively denoted as $\Phi$ and $\Theta$, $U^{i}$ and $S^{i}$, and $M^{ij}$ in the following of this work. As we will see, because of its definition, we have that the condition $U^{i}\!=\!0$ can never occur. However, all other bi-linear quantities can be equal to zero or not: if none of those quantities is equal to zero in general, we talk about \emph{regular} spinor fields, which is essentially the Dirac spinor field, lying at the bases of Quantum Mechanics; if both $\Phi\!=\!\Theta\!=\!0$ we talk instead about \emph{singular} spinor fields, technically known as \emph{flag-dipole} spinor fields, whose mathematical existence has been recently studied in general terms \cite{Vignolo:2011qt, daRocha:2013qhu, Fabbri:2020elt}. If one is not interested in flag-dipole spinor fields in general, one may still find interest in the fact that flag-dipoles can be restricted to two sub-classes according to whether $M^{ij}\!=\!0$ or $S^{i}\!=\!0$ respectively called \emph{dipole} and \emph{flagpole} spinors, and respectively containing the Weyl and Majorana spinors, whose existence is predicted in the Standard Model of Particle Physics.

The flagpoles, algebraically described as Majorana spinors, have acquired peculiar attention in recent times because they are taken as the basis for a new type of spinor field, called Elko, whose dynamical character is that of verifying scalar-like field equations \cite{Ahluwalia:2004ab, Ahluwalia:2004sz, Ahluwalia:2016rwl, Ahluwalia:2016jwz, AH, daRocha:2007pz, HoffdaSilva:2009is, daRocha:2008we, Lee:2014opa, Dvoeglazov:1995eg, Ahluwalia:2019etz}. The rationale for these newly introduced fields is that they must verify second-order differential field equations because they cannot verify first-order differential field equations \cite{Ahluwalia:2019etz}. However, we believe that this assertion is of limited validity. We will show that while specific Elko cannot be solutions of the Dirac equation, general Elko can be solutions of the Dirac equations. The passage from a specific to a more general Elko (or Majorana) spinor will be achieved via an extension of the concept of plane-wave expansion. In turn, this extension will be accomplished by considering the most complete covariant derivative of Majorana (flagpole) spinors.

To do so, we will have spinor fields re-formulated in what is known as polar decomposition. The polar form is the idea of writing a complex function as the product of a module times a unitary phase. When this is done for the wave function, Quantum Mechanics converts into a type of fluid dynamics, where module and unitary phase are linked to the density and the velocity of the matter distribution, while the Schr{\"o}dinger equation splits into a Hamilton-Jacobi and a continuity equation \cite{nonrelB, nonrelT}. In presence of spin the wave function is a spinor with two components, accounting for both helicities, mixing under rotations, and in relativistic cases the wave function is a spinor with four components, accounting for both helicities and both chiralities, mixing under rotations and boosts, so that care must be exercised to implement the polar decomposition while respecting covariance, but a fully general covariant polar re-formulation can be done just as well \cite{jl1, jl2}. As a consequence, Relativistic Quantum Mechanics converts into a generalized type of relativistic fluid dynamics, with the Dirac equations undergoing an analogous polar decomposition \cite{B, T, Fabbri:2023onb}.

From a methodological perspective, our analysis will focus on generality. For spinor fields this is a delicate point because there is a number of ways in which researchers have approached the problem: for quantum field theorists, the spinor field is an expansion in operators verifying certain commutation relations; for mathematical physicists, the spinor is the solution of a differential equation; for mathematicians, it is a section of the spinor bundle. But there is one element that is common to all, that is the spinor field is an object that transforms in a specific way under spinor transformations. This is the only feature of spinor fields that we are going to employ in this work.

The paper is organized as follows: in section \ref{general}, we will recall the basic concepts of the differential geometry of spin manifolds that we will need for the following of the paper; in section \ref{polar}, we will start our study by converting in polar form all spinors of the various Lounesto classes, for both regular and singular spinor fields, for both dipole and flagpole spinors; in section \ref{flagpole}, we will focus on flagpoles, discussing the conditions under which they can be or not solutions to the Dirac equations, especially the role of a possible extension of the plane-wave expansion; in section \ref{Fourier}, we will deepen the discussion on the plane-wave expansion in its enlarged form in a more detailed manner.
\section{General Geometry of Spinor Fields}
\label{general}
As anticipated in the introduction, the only element that we will use to define spinor fields is their transformation law. Therefore, it is essential to start by clarifying as much as possible the character of the possible transformations.

We will work on $(1\!+\!3)$-dimensional manifolds with general curvature. This means that tensor fields will be taken as multi-linear maps defined on the space-time and transforming in terms of general diffeomorphisms. In our convention, they will be denoted with Greek indices. Bases of dual tetrad fields will be assumed to exist, and they will be taken ortho-normal. As a consequence, tetrads $e^{a}_{\nu}$ and $e_{a}^{\nu}$ will have one Greek index, with respect to which they transform in terms of general diffeomorphisms, but also a Latin index, with respect they transform in terms of real Lorentz transformations. If $V^{\sigma}$ is a vector transforming according to diffeomorphisms the object $V^{\sigma}e^{i}_{\sigma}\!=\!V^{i}$ is the same vector transforming according to real Lorentz transformations. This conversion from Greek to Latin indices can be done on every index, in case of generic tensors. Henceforth, we will retain the right to freely convert indices in this manner throughout the work. Clifford matrices $\boldsymbol{\gamma}^{a}$ are defined to be $4\!\times\!4$ complex matrices verifying $\{\boldsymbol{\gamma}^{a},\boldsymbol{\gamma}^{b}\}\!=\!2\mathbb{I}\eta^{ab}$ where $\eta^{ab}$ is the Minkowski matrix. Then $[\boldsymbol{\gamma}^{a},\boldsymbol{\gamma}^{b}]\!=\!4\boldsymbol{\sigma}^{ab}$ are the generators of the complex Lorentz transformations. With the completely antisymmetric pseudo-tensor $\varepsilon_{abcd}$ one can see that $2i\boldsymbol{\sigma}_{ab}\boldsymbol{\pi}\!=\!\varepsilon_{abcd}\boldsymbol{\sigma}^{cd}$ implicitly defining the parity-odd matrix $\boldsymbol{\pi}$ whose existence tells that the complex Lorentz transformations are reducible (this matrix is usually denoted as a gamma with an index five, which has no sense in the space-time, so we prefer to use a notation with no index).

The complex Lorentz transformations are $\boldsymbol{\Lambda}\!=\!\exp{(\frac{1}{2}\theta_{ab}\boldsymbol{\sigma}^{ab})}$ where $\theta_{ij}\!=\!-\theta_{ji}$ are the parameters, which in general can be local. For completeness, we will also consider a gauge transformation $\exp{(iq\theta)}$ where $\theta$ is the parameter and $q$ is a generic constant that will eventually be interpreted as the electric charge. With these two transformations we can define $\boldsymbol{S}\!=\!\boldsymbol{\Lambda}e^{iq\theta}$ as the spinor transformation. This can be written as $\boldsymbol{S}\!=\!\exp{(\frac{1}{2}\theta_{ab}\boldsymbol{\sigma}^{ab}\!+\!iq\theta)}$ in an equivalent manner. We define the spinor fields as what transforms according to
\begin{eqnarray}
&\psi\!\rightarrow\!\boldsymbol{S}\psi\ \ \ \ \ \ \ \ \mathrm{and}
\ \ \ \ \ \ \ \ \overline{\psi}\!\rightarrow\!\overline{\psi}\boldsymbol{S}^{-1}
\label{spin}
\end{eqnarray}
with $\overline{\psi}\!=\!\psi^{\dagger}\boldsymbol{\gamma}^{0}$ as adjoint procedure. With a pair of adjoint spinors we can construct the spinorial bi-linears
\begin{eqnarray}
&\Theta\!=\!i\overline{\psi}\boldsymbol{\pi}\psi\ \ \ \ \ \ \ \ 
\Phi\!=\!\overline{\psi}\psi\\
&S^{a}\!=\!\overline{\psi}\boldsymbol{\gamma}^{a}\boldsymbol{\pi}\psi\ \ \ \ \ \ \ \ 
U^{a}\!=\!\overline{\psi}\boldsymbol{\gamma}^{a}\psi\\
&\Sigma^{ij}\!=\!2\overline{\psi}\boldsymbol{\sigma}^{ij}\boldsymbol{\pi}\psi\ \ \ \ \ \ \ \ 
M^{ij}\!=\!2i\overline{\psi}\boldsymbol{\sigma}^{ij}\psi,
\end{eqnarray}
where the scalars, the vectors and the tensors are found in the first, second and third line, whereas the parity-odd and parity-even tensors are found on the left and right columns, respectively. All these quantities are real. They are tied by the relationships
\begin{eqnarray}
&M_{ik}U^{i}=\Theta S_{k}\ \ \ \ \ \ \ \ \ \ \ \ \ \ \ \ \Sigma_{ik}U^{i}\!=\!\Phi S_{k}\\ 
&M_{ik}S^{i}=\Theta U_{k}\ \ \ \ \ \ \ \ \ \ \ \ \ \ \ \ \Sigma_{ik}S^{i}\!=\!\Phi U_{k}
\end{eqnarray}
\begin{eqnarray}
&M_{ab}\Phi\!-\!\Sigma_{ab}\Theta\!=\!U^{j}S^{k}\varepsilon_{jkab}\ \ \ \ \ \ \ \
M_{ab}\Theta\!+\!\Sigma_{ab}\Phi\!=\!U_{[a}S_{b]}
\end{eqnarray}
\begin{eqnarray}
&\frac{1}{2}M_{ab}M^{ab}\!=\!-\frac{1}{2}\Sigma_{ab}\Sigma^{ab}
\!=\!\Phi^{2}\!-\!\Theta^{2}\label{orthonormal}\\
&\frac{1}{2}M_{ab}\Sigma^{ab}\!=\!-2\Theta\Phi
\end{eqnarray}
\begin{eqnarray}
&U_{a}U^{a}\!=\!-S_{a}S^{a}
\!=\!\Theta^{2}\!+\!\Phi^{2}\label{orthonorm}\\
&U_{a}S^{a}=0
\end{eqnarray}
known as Fierz re-arrangements, and showing that albeit linearly independent these bi-linears are not independent.

From the tetrads we can derive the spin connection, which we will indicate as $C_{ij\mu}$. A gauge potential will also be included for generality, and denoted as $A_{\mu}$. With these elements one can prove that
\begin{eqnarray}
&\boldsymbol{\nabla}_{\mu}\psi\!=\!\partial_{\mu}\psi
\!+\!\frac{1}{2}C_{ij\mu}\boldsymbol{\sigma}^{ij}\psi\!+\!iqA_{\mu}\psi
\end{eqnarray}
is the spinorial covariant derivative of the spinor fields.

Finally, the dynamics will be assigned by the Dirac field equations
\begin{eqnarray}
&i\boldsymbol{\gamma}^{\mu}\boldsymbol{\nabla}_{\mu}\psi\!-\!m\psi\!=\!0\label{D}
\end{eqnarray}
in which $m$ is the mass of the spinor field.

We recall that the Dirac equation is invariant under the discrete transformation
\begin{eqnarray}
&\psi\!\rightarrow\!\boldsymbol{\pi}\psi
\label{discrete}\\
&m\!\rightarrow\!-m
\label{mass}
\end{eqnarray}
which we are going to call $M$ transformation. Another discrete transformation of interest is the charge conjugation
\begin{eqnarray}
&\psi\!\rightarrow\!i\boldsymbol{\gamma}^{2}\psi^{\ast}
\label{C}\\
&q\!\rightarrow\!-q
\label{q}
\end{eqnarray}
up to a phase of no importance, and called for short $C$ transformation.

Some of the concepts recalled in this section have been given only informally because everything said so far is basic knowledge of the differential geometry of spin manifold, found in common textbooks on spinor field theories. It was important however to set the conventions. In the next section \ref{polar} we will examine the Lounesto classes in polar form.
\section{Polar Decomposition of the Lounesto Classes}
\label{polar}
We begin our study of the Lounesto classes by noticing, from \eqref{orthonorm}, that $U_{a}U^{a}\!=\!\Theta^{2}\!+\!\Phi^{2}\!\geqslant\!0$. This allows us to split two major sub-cases: one is when in general not both scalar and pseudo-scalar vanish, and the other is when they are both equal to zero identically. We consider the two cases in the next two sub-sections.
\subsection{Regular Spinor Fields}
For regular spinor fields, not both scalar and pseudo-scalar vanish: hence $U_{a}U^{a}\!>\!0$. This means that the velocity vector $U_{a}$ is time-like. With this information, it is possible to demonstrate that any regular spinor field can always be written, in chiral representation, in the polar form
\begin{eqnarray}
&\psi\!=\!\phi e^{-\frac{i}{2}\beta\boldsymbol{\pi}}
\boldsymbol{L}^{-1}\left(\!\begin{tabular}{c}
$1$\\
$0$\\
$1$\\
$0$
\end{tabular}\!\right)
\label{spinor}
\end{eqnarray}
for some function $\phi$ and $\beta$, called module and chiral angle, and for some matrix $\boldsymbol{L}$, which has the structure of a spinor transformation \cite{jl1, jl2}, \cite{Fabbri:2021mfc}. By inserting the polar form into the bi-linears one can compute
\begin{eqnarray}
&\Theta\!=\!2\phi^{2}\sin{\beta}\ \ \ \ \ \ \ \ \Phi\!=\!2\phi^{2}\cos{\beta}
\end{eqnarray}
showing that the module and chiral angle are a scalar and a pseudo-scalar. The spin axial-vector and velocity vector can be normalized as
\begin{eqnarray}
&S^{a}\!=\!2\phi^{2}s^{a}\ \ \ \ \ \ \ \ U^{a}\!=\!2\phi^{2}u^{a}\label{u}
\end{eqnarray}
verifying the constraints
\begin{eqnarray}
&u_{a}u^{a}\!=\!-s_{a}s^{a}\!=\!1\\
&u_{a}s^{a}\!=\!0
\end{eqnarray}
showing that velocity vector and spin axial-vector have only $3$ independent components each. These are the three rapidities and the three Euler angles encoded as parameters of the $\boldsymbol{L}$ matrix. We have that
\begin{eqnarray}
&\Sigma_{ab}\!=\!2\phi^{2}(\cos{\beta}u_{[a}s_{b]}\!-\!\sin{\beta}u^{j}s^{k}\varepsilon_{jkab})\\
&M_{ab}\!=\!2\phi^{2}(\cos{\beta}u^{j}s^{k}\varepsilon_{jkab}\!+\!\sin{\beta}u_{[a}s_{b]})
\end{eqnarray}
showing that the tensors can be written in terms of all other variables. As a consequence, all remaining Fierz identities listed in the previous section are satisfied identically. When the regular spinor field is converted into polar form, its $4$ complex, or $8$ real, components are re-configured in such a way that the $2$ real scalar degrees of freedom $\phi$ and $\beta$ remain isolated from the $6$ components that can always be transferred into the frame $\boldsymbol{L}$ \cite{Fabbri:2021mfc}.

At the differential level, it is possible to demonstrate that in general
\begin{eqnarray}
&\boldsymbol{L}^{-1}\partial_{\mu}\boldsymbol{L}
\!=\!\frac{1}{2}\partial_{\mu}\xi^{ab}\boldsymbol{\sigma}_{ab}
\!+\!iq\partial_{\mu}\xi\mathbb{I}\label{LdL}
\end{eqnarray}
for some $\xi$ and $\xi^{ab}$ \cite{Fabbri:2021mfc}. Then we can define
\begin{eqnarray}
&\partial_{\mu}\xi_{ij}\!-\!C_{ij\mu}\!\equiv\!R_{ij\mu}\label{R}\\
&q(\partial_{\mu}\xi\!-\!A_{\mu})\!\equiv\!P_{\mu}\label{P}
\end{eqnarray}
which are proven to be a real vector and a real tensor, called gauge and space-time tensorial connections \cite{Fabbri:2021mfc}. Then
\begin{eqnarray}
&\boldsymbol{\nabla}_{\mu}\psi\!=\!(-\frac{i}{2}\nabla_{\mu}\beta\boldsymbol{\pi}
\!+\!\nabla_{\mu}\ln{\phi}\mathbb{I}
\!-\!iP_{\mu}\mathbb{I}\!-\!\frac{1}{2}R_{ij\mu}\boldsymbol{\sigma}^{ij})\psi
\label{decregspinder}
\end{eqnarray}
as the covariant derivative of regular spinor fields in polar form. It tells us that the most complete covariant derivative of spinor fields can always be written as a matrix left-multiplying the spinor field itself. Such a matrix contains two contributions coming from the derivative of the two degrees of freedom $\phi$ and $\beta$ and two contributions coming from the gauge and space-time tensorial connection. It is to be noticed that the tensorial connection $R_{ij\mu}$ plays for space-time transformations the same role that the tensorial connection $P_{\mu}$ plays for gauge transformations. We notice also that if there were only the contribution coming from $P_{\mu}$ expression \eqref{decregspinder} would reduce to $i\boldsymbol{\nabla}_{\mu}\psi\!=\!P_{\mu}\psi$, which is the plane-wave condition of quantum field theory \cite{ps}, and telling us that $P_{\mu}$ is the momentum of the spinor field.

Because the polar form has been extended, by means of the tensorial connections, to the differential structure, we acquire the possibility to express the dynamics in polar variables. So, the Dirac field equations \eqref{D} in polar form are
\begin{eqnarray}
&\nabla_{\mu}\beta\!+\!B_{\mu}\!-\!2P^{\iota}u_{[\iota}s_{\mu]}
\!+\!2ms_{\mu}\cos{\beta}\!=\!0\label{regb}\\
&\nabla_{\mu}\ln{\phi^{2}}\!+\!R_{\mu}\!-\!2P^{\rho}u^{\nu}s^{\alpha}\varepsilon_{\mu\rho\nu\alpha}
\!+\!2ms_{\mu}\sin{\beta}\!=\!0\label{regm}
\end{eqnarray}
where
\begin{eqnarray}
&\frac{1}{2}\varepsilon_{\mu\alpha\pi\iota}R^{\alpha\pi\iota}\!=\!B_{\mu}\label{R1}\\
&R_{\mu\nu\sigma}g^{\nu\sigma}\!=\!R_{\mu}\label{R2}
\end{eqnarray}
were introduced. Equations (\ref{regb}-\ref{regm}) specify all space-time derivatives of both degrees of freedom given by module and chiral angle, and they are equivalent to the initial Dirac field equations \eqref{D} \cite{Fabbri:2023onb}.

The $C$ transformation (\ref{C}-\ref{q}) acts on spinor fields in polar form (\ref{spinor}) according to
\begin{eqnarray}
&i\boldsymbol{\gamma}^{2}\psi^{\ast}\!=\!\phi e^{-\frac{i}{2}\beta\boldsymbol{\pi}}
\boldsymbol{L}^{-1}\left(\!\begin{tabular}{c}
$0$\\
$-1$\\
$0$\\
$1$
\end{tabular}\!\right)\!\equiv\!\phi e^{-\frac{i}{2}(\beta+\pi)\boldsymbol{\pi}}
e^{\frac{i}{2}\pi\mathbb{I}}\boldsymbol{L}^{-1}\left(\!\begin{tabular}{c}
$0$\\
$1$\\
$0$\\
$1$
\end{tabular}\!\right)
\label{Cpolar}
\end{eqnarray}
which is again in polar form, up to a global constant phase, apart from the helicity flip and the shift $\beta\!\rightarrow\!\beta\!+\!\pi$ (the parity-odd matrix $\boldsymbol{\pi}$ and the constant $\pi$ are denoted by the same letter in bold-face and in normal style, so there should be no confusion here). The reading of \eqref{Cpolar} is immediate: it tells us that the $C$ transformation is just the $M$ transformation complemented by the helicity flip. In the standard understanding of quantum field theory, the $C$ transformation is taken to implement the passage from matter to antimatter, with the problem that, as well known, a $C$ transformation also inverts the sign of the energy. In quantum field theory this problem is addressed by second-quantization of fields. However, this problem would not even arise if there were no helicity flip. This is what the $M$ transformation does. The $M$ transformation, being the $C$ transformation with no helicity flip, may implement the matter/antimatter duality without the energy problem, as first pointed out in \cite{bz}.
\subsection{Singular Spinor Fields}
For singular spinor fields, both scalar and pseudo-scalar vanish identically: thus $U_{a}U^{a}\!\equiv\!0$. This means that the velocity vector $U_{a}$ is light-like. Moreover, because of \eqref{orthonormal}, it also means that $M_{ab}M^{ab}\!\equiv\!0$. And this means that the space-space and space-time components of $M_{ab}$ have the same norm and they are orthogonal to each other. With all these elements, it is possible to demonstrate that any singular spinor field, or flag-dipole spinor field, can always be written, in chiral representation, in the polar form
\begin{eqnarray}
&\psi\!=\!\frac{1}{\sqrt{2}}(\mathbb{I}\cos{\frac{\alpha}{2}}
\!-\!\boldsymbol{\pi}\sin{\frac{\alpha}{2}})\boldsymbol{L}^{-1}\left(\!\begin{tabular}{c}
$1$\\
$0$\\
$0$\\
$1$
\end{tabular}\!\right)
\label{flagdipole}
\end{eqnarray}
for some function $\alpha$ and some matrix $\boldsymbol{L}$ \cite{Fabbri:2021mfc}. By inserting the polar form into the bi-linears one can compute
\begin{eqnarray}
&S^{a}\!=\!-\sin{\alpha}U^{a}
\end{eqnarray}
showing that $\alpha$ is a pseudo-scalar field. All Fierz identities are shown to trivialize.

Still with (\ref{R}-\ref{P}) we can write
\begin{eqnarray}
&\boldsymbol{\nabla}_{\mu}\psi\!=\!(-\frac{1}{2}\tan{\alpha}\nabla_{\mu}\alpha\mathbb{I}
\!-\!\frac{1}{2}\sec{\alpha}\nabla_{\mu}\alpha\boldsymbol{\pi}
\!-\!iP_{\mu}\mathbb{I}\!-\!\frac{1}{2}R_{ij\mu}\boldsymbol{\sigma}^{ij})\psi
\label{decsingspinder}
\end{eqnarray}
as the covariant derivative of singular spinor fields in polar form.

And with the same (\ref{R1}-\ref{R2}) we have
\begin{eqnarray}
&(\varepsilon^{\mu\rho\sigma\nu}\nabla_{\mu}\alpha\sec{\alpha}
\!-\!2P^{[\rho}g^{\sigma]\nu})M_{\rho\sigma}\!=\!0\label{sing1}\ \ \ \
\ \ \ \ [-B^{\sigma}\varepsilon_{\sigma\mu\rho\nu}\!+\!R_{[\mu}g_{\rho]\nu}
\!+\!g_{\nu[\mu}\nabla_{\rho]}\alpha\tan{\alpha}]
M_{\eta\zeta}\varepsilon^{\mu\rho\eta\zeta}\!=\!0\label{sing2}\\
&M_{\rho\sigma}(g^{\nu[\rho}\nabla^{\sigma]}\alpha\sec{\alpha}
\!-\!2P_{\mu}\varepsilon^{\mu\rho\sigma\nu})\!+\!4m\sin{\alpha}U^{\nu}\!=\!0\label{sing3}\ \ \ \
\ \ \ \ [-B^{\sigma}\varepsilon_{\sigma\mu\rho\nu}\!+\!R_{[\mu}g_{\rho]\nu}
\!+\!g_{\nu[\mu}\nabla_{\rho]}\alpha\tan{\alpha}]M^{\mu\rho}\!+\!4mU_{\nu}\!=\!0\label{sing4}
\end{eqnarray}
as the Dirac field equations in polar form. Equations (\ref{sing1}-\ref{sing4}) are equivalent to the original Dirac field equations \cite{Fabbri:2021mfc}.

The $M$ transformation works as above. Also $C$ works as above but with the addition that $\alpha\!\rightarrow\!-\alpha$. It is important to wonder about the significance of the singular spinor field. From a mathematical perspective, such a flag-dipole is a well-defined spinor field, with a system of field equations that determine all space-time derivatives of the degree of freedom, and admitting solutions \cite{Vignolo:2011qt, daRocha:2013qhu}. Yet, we have no evidence for this kind of behaviour anywhere in nature.

Another peculiar feature will be described after the introduction of dipole and flagpole spinors.

\subsubsection{Dipole Spinors}
With the additional restriction $M^{ab}\!\equiv\!0$ flag-dipole spinor fields restrict to dipole spinors. It is easy to show that condition $M^{ab}\!=\!0$ is equivalent to $\alpha\!=\!\pm\pi/2$ and therefore
\begin{eqnarray}
&\psi\!=\!\boldsymbol{L}^{-1}\left(\!\begin{tabular}{c}
$1$\\
$0$\\
$0$\\
$0$
\end{tabular}\!\right)\ \ \ \ \ \ \ \ \mathrm{or}\ \ \ \ \ \ \ \ 
\psi\!=\!\boldsymbol{L}^{-1}\left(\!\begin{tabular}{c}
$0$\\
$0$\\
$0$\\
$1$
\end{tabular}\!\right),
\label{Weyl}
\end{eqnarray}
which are left-handed and right-handed Weyl spinors. The converse is also true. So, the class of dipoles is exhausted completely by the Weyl spinors \cite{Fabbri:2021mfc}. For left/right-handed Weyl spinors we have $S^{a}\!=\!\mp U^{a}$ respectively.

The covariant derivative in polar form reduces to
\begin{eqnarray}
&\boldsymbol{\nabla}_{\mu}\psi
\!=\!(-iP_{\mu}\mathbb{I}\!-\!\frac{1}{2}R_{ij\mu}\boldsymbol{\sigma}^{ij})\psi\label{derW}.
\end{eqnarray}

As for the dynamics, dipoles are forced to have $m\!=\!0$ and thus their Dirac field equations are
\begin{eqnarray}
&R_{\mu}U^{\mu}\!=\!0\ \ \ \
\ \ \ \ (-B_{\mu}\!\pm\!2P_{\mu})U^{\mu}\!=\!0\ \ \ \
\ \ \ \ [(-B_{\mu}\!\pm\!2P_{\mu})\varepsilon^{\mu\rho\alpha\nu}
\!+\!g^{\rho[\alpha}R^{\nu]}]U_{\rho}\!=\!0
\end{eqnarray}
respectively for the left/right-handed case. Weyl spinors are eigenstates of the $M$ transformation.

\subsubsection{Flagpole Spinors}
With the additional restriction $S^{a}\!\equiv\!0$ flag-dipole spinor fields restrict to flagpole spinors. Condition $S^{a}\!=\!0$ is equivalent to $\alpha\!=\!0$ or $\alpha\!=\!\pi$ and consequently
\begin{eqnarray}
&\psi\!=\!\frac{1}{\sqrt{2}}\boldsymbol{L}^{-1}\left(\!\begin{tabular}{c}
$1$\\
$0$\\
$0$\\
$1$
\end{tabular}\!\right)\ \ \ \ \ \ \ \ \mathrm{or}\ \ \ \ \ \ \ \
\psi\!=\!\frac{1}{\sqrt{2}}\boldsymbol{L}^{-1}\left(\!\begin{tabular}{c}
$1$\\
$0$\\
$0$\\
$-1$
\end{tabular}\!\right),
\label{Majorana}
\end{eqnarray}
as self-conjugated and antiself-conjugated Majorana spinors. The converse is true too. Thus, the class of flagpoles is exhausted by the Majorana spinors \cite{Fabbri:2021mfc}. The Majorana and anti-Majorana spinors are such that $S^{a}\!=\!0$.

The covariant derivative in polar form reduces to
\begin{eqnarray}
&\boldsymbol{\nabla}_{\mu}\psi\!=\!-\frac{1}{2}R_{ij\mu}\boldsymbol{\sigma}^{ij}\psi
\label{derM}
\end{eqnarray}
due to the fact that $P_{\mu}\!=\!0$ for flagpoles. To see this, consider that self-conjugated or antiself-conjugated of $C$ can be explicitly written as $\psi\!=\!i\boldsymbol{\gamma}^{2}\psi^{\ast}$ up to a phase. Then $\boldsymbol{\nabla}^{\mu}\psi\!=\!i\boldsymbol{\gamma}^{2}\boldsymbol{\nabla}^{\mu}\psi^{\ast}$. Requiring this condition on \eqref{decsingspinder} when $\alpha\!=\!0$ or $\alpha\!=\!\pi$ implies that $(-iP^{\mu}\mathbb{I}
\!-\!\frac{1}{2}R^{ij\mu}\boldsymbol{\sigma}_{ij})\psi
\!=\!i\boldsymbol{\gamma}^{2}(iP^{\mu}\mathbb{I}
\!-\!\frac{1}{2}R^{ij\mu}\boldsymbol{\sigma}_{ij}^{\ast})\psi^{\ast}
\!=\!(iP^{\mu}\mathbb{I}\!-\!\frac{1}{2}R^{ij\mu}\boldsymbol{\sigma}_{ij})\psi$ and hence $P_{\mu}\!=\!0$ identically.

As for the dynamics, flagpoles have field equations
\begin{eqnarray}
&R_{\mu}U^{\mu}\!=\!0\ \ \ \
\ \ \ \ B_{\mu}U^{\mu}\!=\!0\ \ \ \
\ \ \ \ (-B_{\mu}\varepsilon^{\mu\rho\alpha\nu}
\!+\!g^{\rho[\alpha}R^{\nu]})U_{\rho}\!+\!2m\!M^{\alpha\nu}\!\!=\!0\label{origano}
\end{eqnarray}
in general. As already stated, Majorana spinors are eigenstates of the $C$ transformation, and so $q\!=\!0$.

We are now in position to discuss the other peculiar feature of flag-dipoles, anticipated at the beginning of this section. As noticed, for flag-dipole spinor fields the degree of freedom $\alpha$ is determined as solutions of the Dirac field equations. Then, $\alpha$ it is to be considered as a physical field. As such, it can be free to vary between different values, in particular those for which $\alpha\!=\!0$ and $\alpha\!=\!\pi/2$. What this means is that in general we should be able to have the transmutation of a (neutral) dipole into a (massless) flagpole. This occurrence, however, has never been observed.

The last case we considered, the flagpole, or Majorana spinor, is interesting because in recent times they have been taken into account to define new types of spinor fields, called Elko, characterized as being solutions of second-order field equations, since they cannot be solutions of first-order field equations, as discussed in \cite{Ahluwalia:2004ab, Ahluwalia:2004sz, Ahluwalia:2016rwl, Ahluwalia:2016jwz, AH} and thoroughly reviewed in \cite{Ahluwalia:2019etz}. However, we now have all elements needed to assess the matter in the most general situation.
\section{Majorana Spinors and Tensorial Connections}
\label{flagpole}

\subsection{The Elko spinor}
The last section ended with the introduction of Majorana spinors. They were characterized as being eigenstates of $C$ conjugation, and proven to have $q\!\equiv\!0$ as well as $P_{i}\!\equiv\!0$. Their Dirac field equations in polar form were presented.

The known problem that Majorana spinors have in quantum field theory is that there does not seem to be any way to write a Majorana Lagrangian, since this would require terms like $\overline{\psi}\psi$ to appear, and we have seen that these terms are zero for flagpoles, and more in general flag-dipoles. In Quantum Field Theory, the issue is solved by having Majorana spinors re-interpreted as Grassmann-valued fields. However, such Grassmannization is a procedure that does not come without some criticism (see \cite{Ahluwalia:2019etz} and references therein). In alternative to Grassmannization, one can consider a new type of Majorana spinor, called Elko spinor, in terms of which the aforementioned problem is overcome with a new definition of dual \cite{Ahluwalia:2004ab, Ahluwalia:2004sz, Ahluwalia:2016rwl, Ahluwalia:2016jwz, AH}. An interesting consequence of Elko spinors is that they must solve second-order field equations \cite{Lee:2014opa, Dvoeglazov:1995eg}. The reason is that they are not supposed to solve first-order field equations \cite{Ahluwalia:2019etz}.

To see why, let us recall some general definitions. From now on, flagpoles, or Majorana spinors, as well as Elko spinors, will be designated by $\lambda$. Charge-conjugation can be implemented either as
\begin{eqnarray}
&C\lambda^{S}\!=\!+\lambda^{S}\ \ \ \ \ \ \ \ \mathrm{or\ as} \ \ \ \ \ \ \ \ C\lambda^{A}\!=\!-\lambda^{A}
\end{eqnarray}
with $S$ and $A$ standing for the self- and the antiself-conjugated Elko. Explicitly we have
\begin{eqnarray}
&\lambda^{S}_{+}\!=\!\chi\left(\!\begin{tabular}{c}
$0$\\
$-e^{-i\omega}$\\
$e^{i\omega}$\\
$0$
\end{tabular}\!\right)\ \ \ \ \ \ \ \ \mathrm{and} \ \ \ \ \ \ \ \
\lambda^{A}_{+}\!=\!\chi\left(\!\begin{tabular}{c}
$0$\\
$e^{-i\omega}$\\
$e^{i\omega}$\\
$0$
\end{tabular}\!\right)\label{+}\\
&\lambda^{S}_{-}\!=\!\chi\left(\!\begin{tabular}{c}
$e^{-i\omega}$\\
$0$\\
$0$\\
$e^{i\omega}$
\end{tabular}\!\right)\ \ \ \ \ \ \ \ \mathrm{and} \ \ \ \ \ \ \ \
\lambda^{A}_{-}\!=\!\chi\left(\!\begin{tabular}{c}
$-e^{-i\omega}$\\
$0$\\
$0$\\
$e^{i\omega}$
\end{tabular}\!\right)\label{-}
\end{eqnarray}
where the plus and minus signs account for helicity-up and helicity-down states. Nevertheless, we have to notice that, even in a theory in which fields are local, when the spinor transformation is point-dependent one can always employ a suitable transformation to trivialize both functions inside the Majorana and Elko fields. Therefore, we remain with
\begin{eqnarray}
&\lambda^{S}_{+}\!=\!\left(\!\begin{tabular}{c}
$0$\\
$-1$\\
$1$\\
$0$
\end{tabular}\!\right)\ \ \ \ \ \ \ \ \mathrm{and} \ \ \ \ \ \ \ \
\lambda^{A}_{+}\!=\!\left(\!\begin{tabular}{c}
$0$\\
$1$\\
$1$\\
$0$
\end{tabular}\!\right)\label{+c},\\
&\lambda^{S}_{-}\!=\!\left(\!\begin{tabular}{c}
$1$\\
$0$\\
$0$\\
$1$
\end{tabular}\!\right)\ \ \ \ \ \ \ \ \mathrm{and} \ \ \ \ \ \ \ \
\lambda^{A}_{-}\!=\!\left(\!\begin{tabular}{c}
$-1$\\
$0$\\
$0$\\
$1$
\end{tabular}\!\right)\label{-c}.
\end{eqnarray}
This fact can be seen by noticing that for (\ref{+}-\ref{-}), a boost along the third axis acts as a scaling while a rotation around the third axis acts as a chiral phase. Alternatively, one may simply consider that, by re-expressing flagpoles in polar form (\ref{Majorana}), the Lorentz transformation $\boldsymbol{L}$ is what would take (\ref{+}-\ref{-}) into (\ref{+c}-\ref{-c}) up to a constant phase.

Remark that $\lambda^{S}_{+}\!=\!\boldsymbol{\pi}\lambda^{A}_{+}$ and $\lambda^{S}_{-}\!=\!\boldsymbol{\pi}\lambda^{A}_{-}$ and that the above (\ref{+c}-\ref{-c}) are the helicity-flip of each other. Consequently, if we can find a time-like vector $v^{\mu}$, defining $p^{\mu}\!=\!mv^{\mu}$ we can write the covariant expressions
\begin{eqnarray}
&\boldsymbol{\gamma}_{\mu}p^{\mu}\lambda^{S}_{+}\!+\!m\lambda^{A}_{-}=0,\ \ \ \
\ \ \ \ \boldsymbol{\gamma}_{\mu}p^{\mu}\lambda^{S}_{-}\!-\!m\lambda^{A}_{+}=0,\ \ \ \
\ \ \ \ \boldsymbol{\gamma}_{\mu}p^{\mu}\lambda^{A}_{+}\!-\!m\lambda^{S}_{-}=0,\ \ \ \
\ \ \ \ \boldsymbol{\gamma}_{\mu}p^{\mu}\lambda^{A}_{-}\!+\!m\lambda^{S}_{+}=0,\label{Elkos}
\end{eqnarray}
as straightforward to see. In fact, take for instance the first. By writing it as $\boldsymbol{\gamma}_{\mu}v^{\mu}\lambda^{S}_{+}\!+\!\lambda^{A}_{-}=0$ we can boost in the frame where the time-like vector $v^{\mu}$ has only its time component, so that $\boldsymbol{\gamma}_{0}\lambda^{S}_{+}\!+\!\lambda^{A}_{-}=0$, which is verified by (\ref{+c}-\ref{-c}).

This means that the operator $\boldsymbol{\gamma}_{\mu}p^{\mu}\!\pm\!m$ is not annihilated by Elko, so Elko cannot verify Dirac field equations \cite{Ahluwalia:2019etz}.

Is this fact general? Or is its validity limited by hidden assumptions? The answer is that the momentum appearing in \eqref{Elkos} is the kinematic momentum $p^{\mu}\!=\!mv^{\mu}$, defined in terms of the time-like vector $v^{\mu}$. Still, this does not mean that $v^{\mu}$ is the velocity vector $u^{\mu}$ defined in (\ref{u}). And even if we chose $v^{\mu}\!\equiv\!u^{\mu}$, this would not imply that $p^{\mu}\!=\!mu^{\mu}$ is the dynamic momentum $P^{\mu}$ defined in (\ref{P}). This fact is trivial, since we have demonstrated in the previous section that the dynamic momentum $P^{\mu}$ is equal to zero identically for flagpoles in general. As a consequence, it is true that the Dirac operator is not annihilated by Elko. But to a more careful analysis, we can only say that the Dirac operator restricted to have kinematic momentum is not annihilated by Elko. However, we may not be at liberty to choose the restriction to the kinematic momentum. In fact, we may not even be free to choose any restriction to the momentum space since, again, the momentum of Elko is demonstrated to be zero, as already discussed. Because the choice of the kinematic momentum is arbitrary and the dynamic momentum is zero, the wisest behaviour we can have now would be to make no assumption at all and let the Dirac operator be written in its most general expression.
\subsection{The space-time tensorial connection}
The most general covariant derivative of flagpoles in polar form is \eqref{derM}. So the Dirac operator for Elko is
\begin{eqnarray}
&i\boldsymbol{\gamma}^{\mu}\boldsymbol{\nabla}_{\mu}\lambda
\!=\!\frac{1}{2}(iR_{k}\boldsymbol{\gamma}^{k}
\!+\!B_{k}\boldsymbol{\gamma}^{k}\boldsymbol{\pi})\lambda
\end{eqnarray}
and the Dirac equations \eqref{D} become
\begin{eqnarray}
&(iR_{a}\boldsymbol{\gamma}^{a}\!+\!B_{a}\boldsymbol{\gamma}^{a}\boldsymbol{\pi}
\!-\!2m)\lambda\!=\!0.
\label{Dirac}
\end{eqnarray}
Of these we should ask whether there can be solutions in the form of Elko.

For this, notice that under $C$-conjugation the Dirac equation \eqref{Dirac} is invariant. Hence, there should be no obstruction to finding Elko solutions for the Dirac equation \eqref{Dirac}.

An example of such solution, in the simplest possible case, can indeed be found explicitly, for a space-time tensorial connection having a single non-zero component given by
\begin{eqnarray}
&R_{211}\!=\!-2m:
\label{example}
\end{eqnarray}
this would give $R_{2}\!=\!2m$ and the ensuing Dirac equation \eqref{Dirac} would be
\begin{eqnarray}
&\left(\begin{array}{cccc}
\!1 \!&\! 0 \!&\! 0 \!&\! -1\!\\
\!0 \!&\! 1 \!&\! 1 \!&\! 0\!\\
\!0 \!&\! 1 \!&\! 1 \!&\! 0\!\\
\!-1 \!&\! 0 \!&\! 0 \!&\! 1\!\\
\end{array}\right)\lambda\!=\!0
\end{eqnarray}
which has indeed solutions in the form of Elko. In fact, it is easy to check that
\begin{equation}
\lambda\!=\!\left(\!\begin{tabular}{cccc}
$1$\\
$0$\\
$0$\\
$1$
\end{tabular}\!\right),
\end{equation}
which is just $\lambda^{S}_{-}$ in \eqref{-c}, solves it. Less simple solutions of the Dirac equations can be found by employing the polar form: two solutions for the Dirac polar equations \eqref{origano} have been presented in full detail in reference \cite{Fabbri:2023kjy}.

The solution given above for \eqref{example} with the two solutions in \cite{Fabbri:2023kjy} are three counter-examples to the assertion that Elko cannot be solutions of Dirac equations. The Dirac operator does not have any obstruction, and indeed solutions can be found. We will now proceed to analyze these solutions from a constructive perspective.
\section{Doubly-Chiral Plane-Wave Expansion}
\label{Fourier}
In the previous section, the claim that Elko cannot be solutions to the Dirac equations was seen to have a validity limited by too restrictive assumptions on the momentum. And we have seen that when no such assumption was made, general Dirac equations did have solutions in the form of Elko, presenting or discussing three such solutions.

In this last section, we would like to be constructive, analyzing the reasons behind the fact that flagpoles, despite being characterized by the condition $P_{\mu}\!\equiv\!0$, can still solve the Dirac equations.

A way to re-formulate this question is: even when $P_{\mu}\!\equiv\!0$, can we still talk about \emph{some} type of momentum?

Let us consider again the explicit solution seen at the end of the last section. It is given for \eqref{example} by a flagpole with the structure $\lambda^{S}_{-}$ in \eqref{-}. Condition \eqref{derM} can be formally integrated as
\begin{eqnarray}
\lambda=\exp{\left(-\frac{1}{2}\int R_{ij\mu}dx^{\mu}\boldsymbol{\sigma}^{ij}\right)}\lambda_{0}
\label{dcFspinorspecial}
\end{eqnarray}
with $\lambda_{0}$ the constant flagpole given by $\lambda^{S}_{-}$ in \eqref{-c}. After computations, we find
\begin{eqnarray}
\!\!\!\!\!\!\!\!\lambda\!=\!\exp{\left(\!-\frac{1}{2}\!\int\!\!R_{ij\mu}dz^{\mu}
\boldsymbol{\sigma}^{ij}\right)}\lambda_{0}
\!=\!\exp{\left(mz\boldsymbol{\gamma}^{2}\boldsymbol{\gamma}^{1}\right)}\lambda_{0}
\!=\!\left(\begin{array}{cccc}
\!e^{imz} \!&\! 0 \!&\! 0 \!&\! 0\!\\
\!0 \!&\! e^{-imz} \!&\! 0 \!&\! 0\!\\
\!0 \!&\! 0 \!&\! e^{imz} \!&\! 0\!\\
\!0 \!&\! 0 \!&\! 0 \!&\! e^{-imz}\!\\
\end{array}\right)\!\cdot\!\left(\!\begin{tabular}{c}
$1$\\
$0$\\
$0$\\
$1$
\end{tabular}\!\right)\!=\!\left(\begin{tabular}{c}
$e^{imz}$\\
$0$\\
$0$\\
$e^{-imz}$
\end{tabular}\right)
\end{eqnarray}
where $z$ is the coordinate along the axis of propagation. This is just $\lambda^{S}_{-}$ in \eqref{-}. Such a solution splits in two as
\begin{eqnarray}
\!\!\!\!\!\!\!\!\lambda\!=\!\left(\begin{tabular}{c}
$e^{imz}$\\
$0$\\
$0$\\
$e^{-imz}$
\end{tabular}\right)\!=\!e^{imz}\left(\begin{tabular}{c}
$1$\\
$0$\\
$0$\\
$0$
\end{tabular}\right)\!+\!e^{-imz}\left(\begin{tabular}{c}
$0$\\
$0$\\
$0$\\
$1$
\end{tabular}\right),
\end{eqnarray}
showing that it can be seen as a superposition of two dipoles, each of them having a plane-wave coefficient. However, as one can observe, the two coefficients are the inverse of one another.

This already seems to point out the difference with respect to the standard treatment: while with $P_{\mu}$ we have one momentum alone for both chiral parts, with $R_{ij\mu}$ we have something that has the features of a momentum but which can accommodate opposite behaviours for the two chiral parts.

To see this more clearly, let us go back to the general case of spinors $\psi$, and let us consider the polar form of the covariant derivatives (\ref{decregspinder} ,\ref{decsingspinder}, \ref{derW} ,\ref{derM}). For simplicity, let us drop the terms related to the degrees of freedom and focus only on the tensorial connections. The relevant expression, common to all of the above, is
\begin{eqnarray}
\boldsymbol{\nabla}_{\mu}\psi
\!=\!(-iP_{\mu}\mathbb{I}\!-\!\frac{1}{2}R_{ij\mu}\boldsymbol{\sigma}^{ij})\psi\label{der}
\end{eqnarray}
where of course $P_{\mu}\!\equiv\!0$ represents the flagpole. Its formal integration is
\begin{eqnarray}
\psi\!=\!\exp{\left[-\!\int\!\left(iP_{\mu}
\!+\!\frac{1}{2}\boldsymbol{\sigma}^{ij}R_{ij\mu}\right)dx^{\mu}\right]}\psi_{0}
\label{dcFspinor}
\end{eqnarray}
where $\psi_{0}$ is a constant spinor. In absence of $R_{ij\mu}$ this reduces to the usual plane-wave expansion. However, in general, the term in $R_{ij\mu}$, entering in front of a generator $\boldsymbol{\sigma}^{ij}$, is sensible to differences between the two chiral parts. Differences in behaviour between the two chiral parts cannot be seen in non-relativistic cases, where the spinor 'small'-component vanishes, and so the two chiral parts are by construction identical. Nor they can be seen for Weyl spinors, where one of the two chiral parts is missing. However, Majorana spinors have two chiralities, and because the two chiral parts have opposite helicities, they necessarily behave differently. The flag-dipole spinors, which are the most general singular spinors, have two chiral parts whose different behaviour is encoded by $\alpha$. And regular spinors have two chiral parts whose differences are codified by the chiral angle $\beta$. Spinors with non-zero $\beta$ are known to exist. For example, relativistic solutions of the Dirac equation with Coulomb potential, the Hydrogen atom, have a non-null chiral angle.

The discussed extension to a doubly-chiral plane-wave expansion may have considerable importance in all quantum field theoretical computations in which the two chiral parts might display a different behaviour. And it is necessary for flagpoles, since in this case the two chiral parts always display a different behaviour.
\section{Conclusion}
We have reviewed the Lounesto classification in the light of the polar re-formulation, showing the polar decomposition of spinors in all Lounesto classes. After introducing the pair of tensorial connections, we have given the covariant derivative of spinors in all these cases. We have presented the Dirac equations in polar form for each.

The special case of flagpoles, that is the Majorana spinors, was deepened with considerations revolving around the recently introduced Elko \cite{Ahluwalia:2019etz}. The fundamental claim that Elko cannot solve the Dirac equation was criticised on the basis of a too restricted definition of momentum, and we have demonstrated that when no restriction is imposed then Majorana spinors are indeed found to be solutions of the first-order differential Dirac equations.

This fact was then taken as basis for a more constructive idea. By asking what could play the role of the momentum, we have seen that the space-time tensorial connection $R_{ij\mu}$ has some of the features of a momentum and in addition it can encompass the possibly different behaviours of the two chiral projections. Expression \eqref{dcFspinor} represents the usual plane-wave expansion extended so to allow the possibility that the two chiral parts may have opposite momenta.

This doubly-chiral plane-wave expansion, necessary in the case of flagpole spinors, is needed also in all those cases in which the chiral parts have non-trivial differences, as it generally happens for singular and regular spinor fields.

\

{\bf Funding information and additional acknowledgements}. This work has been carried out in the framework of activities of the INFN Research Project QGSKY. This work has been funded by Next Generation EU through the project ``Geometrical and Topological effects on Quantum Matter (GeTOnQuaM)''.

\

{\bf Conflict of interest}. There is no conflict of interest.

\end{document}